\documentclass[showpacs,preprintnumbers,amsmath,amssymb]{revtex4}
\usepackage{graphicx}
\usepackage{dcolumn}
\usepackage{bm}
\begin{document}
\title{Role of Faddeevian anomaly in the s-wave scattering of chiral fermion off dilaton
black holes towards preservation of information}


\author{Anisur Rahaman}
\email{1. anisur.rahman@saha.ac.in, 2. manisurn@gmail.com}
\affiliation{Hooghly Mohsin College, Chinsurah, Hooghly - 712101,
West Bengal, India}

\date{\today}

\begin{abstract}
It was found that s-wave scattering of chiral fermion off dilaton
black-hole when studied with a model generated from the chiral
Schwinger model with standard Jackiw-Rajaraman type of anomaly
provided information non preserving result. However this
scattering problem when studied with a model generated from chiral
Schwinger model with generalized Faddeevian type of anomaly
 rendered information preserving result and  it had well
agreement with Hawking's revised proposal related to information
loss. A minute and equitable investigation in detail has been
carried out here to show how Faddeevian type of anomaly scores
over the Jackiw-Rajaraman type of anomaly in connection with the
s-wave scattering of chiral fermion of dilaton black-hole related
to preservation of information.

\end{abstract}
 \maketitle

Free chiral boson can be thought of in terms of chiral fermion in
$(1+1)$ dimension which is a basic ingredient of heterotic string
theory. So the theories involving chiral boson is interesting in
its own right. There are quite a good mumbler of interesting field
theoretical models involving free as well as interacting chiral
boson \cite{SIG, SONI,BEL,FJ,KH}. So scattering of chiral fermion
off dilaton black-hole would certainly be of interest and the
model presented in the article \cite{STRO} is so beautifully
formulated that s-wave scattering of chiral fermion too can be
studied in a significant manner. In the article \cite{MIT}, the
authors first made an attempt to study the s-wave scattering of
chiral fermion off dilaton black-hole, but they obtained a result
which went against preservation of information. The interacting
(gauged) chiral boson (fermion) part in that case was taken from
the interacting model invented by Harada in \cite{KH}. This model
was generated from the usual chiral Schwinger model with the
standard one para meter class of anomaly introduced by
Jackiw-Rajaraman \cite{JR} under the imposition of a chiral
constraint by hand in the phase space of the theory. Recently, it
is shown in the article \cite{ARCH}, that it can be obtained
directly from the gauged action of Siegel type chiral boson
\cite{SIG, BEL} too.

Although in the article \cite{MIT}, the s-wave scattering of
chiral fermion off dilaton black-hole rendered a result with
nonpreservation of information, in  our article \cite{AR1}, we
have shown that s-wave scattering of chiral fermion with
generalized Faddeevian anomaly \cite{FADDEEV}  leads to an
information preserving result. This information preserving result
related to information loss scenario of course is a gift of the
Faddeevian anomaly and it indeed  agrees with Hawking's revised
suggestion \cite{HAW1}. In this article, an attempt has,
therefore, been made to carry out a minute and detailed
investigation how Faddeevian anomaly changes the scenario related
to information loss in the scattering of chiral fermion off
dilaton black-hole.

In this context, and for the sake of the readers benefit, we
should mention a bit history related to information loss paradox.
It is universally accepted that the issue that received special
attention related to black-hole physics is the possible
information loss when a black-hole evaporates through the form of
thermal Hawking radiation. However from that very time of the
discovery of this thermal radiation by Hawking \cite{HAW}, a
controversy was initiated concerning his proposal related to
information loss \cite{HAW}, since it carried an indication of a
new level of unpredictability within the quantum mechanics by the
gravity. After four decades, Hawking himself moved away from his
own initial proposal and suggested that quantum gravity
interaction did not lead to any loss of information and quantum
coherence did maintain \cite{HAW1}. His revised suggestion on this
issue although gave back a moderately pleasant stratus related to
information loss puzzle, it would be fair to admit that the
information loss related enigma would not yet been well settled
down from all corners. Some investigations are still standing with
information non-preserving result without having any plausible
suggestion towards solvation of the puzzle.

The paper is organized as follows. Sec. II contains a brief
discussion of the model used for studying the s-wave scattering of
fermion off dilaton black-hole. Sec III is devoted to describe the
s-wave scattering of chiral  fermion with the Faddeevian anomaly
with the specialized form developed by Mitra in the article
\cite{PM} with the plausible explanation: why does it fail to
preserve information? In Sec. IV, it is shown how the important
and useful technique, e.g. imposition of chiral constraint
developed by Harada \cite{KH} can miraculously solve the
information loss problem. An over all discussion and conclusion is
contained in Sec V.

\section{{\bf A brief review of the model used for studying s-wave
 scattering of fermion off dilaton black-hole}}
The backscattering of s-wave fermion with magnetically charged-Q
dilaton black-hole is described by the action  \cite{GARF, GIB1,
GIB2, STRO, REV}
\begin{equation}
S_{AF} = \int d^4 x\sqrt{g}[e^{-2\Phi}[R + 4(\nabla\Phi)^2 -
\frac{1}{4}F^2] + i\bar\Psi D\!\!\!/\Psi], \label{PACT}
\end{equation}
The fields $\Psi$  and $\Phi$ represent the charged fermion and
the dilaton field respectively. The factor $\sqrt{g}$ is
associated with the metric tensor $g_{\mu\nu}$, and $F_{\mu\nu}$
is the electromagnetic field strength. The geometry of this
$(3+1)$ dimensional black-hole is constituted with three important
regions: a symmetrical throat region far from the black-hole. As
long as it proceeds near the black-hole the curvature begins to
rise and finally enters into the mouth region. The metric is
approximated to the flat $(1+1)$ dimensional Minkowsky space {\it
times} the round metric on the two sphere with radius $|Q|$ far
along the throat region.
 At
large scale relative to the radius $|Q|$ the following action
therefore results in
\begin{equation}
S_{AF} = \int d^2 x\sqrt{g}[e^{-2\Phi}[R + 4(\nabla\Phi)^2 +
\frac{1}{Q^2} - \frac{1}{4}F^2] + i\bar\Psi D\!\!\!/\Psi],
\label{ACT}
\end{equation}
This action was obtained treating the throat region of $(3+1)$
dimensional black-hole as a compactified form of $(3+1)$ dimension
to $(1+1)$ dimension. The black-hole solution of which was
provided for in the article \cite{GAS}. It is known that in the
extremal limit the geometry is non singular having no horizon, but
when a low energy particle is thrown into that; it acquires a
singularity along with an event horizon \cite{STRO, REV}. We
should  mention at this stage  that following the review
 article \cite{REV} the diatonic equation of motion computed from
 the action
(\ref{ACT}) reads
\begin{equation}
e^{-2\Phi}[R + 4\lambda^2 + 4\nabla^2\Phi -4(\nabla\Phi)^2 -
\frac{1}{4}F_{\mu\nu}F^{\mu\nu}]=0,
\end{equation}
where  $\lambda^2 = \frac{1}{4Q^2}$.  However the linear diatonic
vacuum which is characterized by
\begin{equation}
R= \nabla^2 \Phi = 0, ~~~~(\nabla \Phi)^2 = \lambda^2.
\end{equation}
So we can  introduce a coordinate system $(\tilde{x}, \tau )$ such
that $g_{\mu\nu} = \eta_{\mu \nu}$ and $\Phi= -\lambda \tilde{x}$
in the vacuum. So the square of the line element gets simplified
to $ds^2 = \eta_{\mu\nu}d\tilde{x}^\mu d\tilde{x}^\nu$. Certainly,
$\tau$ stands for $\tilde{x}^0$. The natural dilatonic coupling
for this theory is $e^{\Phi(\tilde{x})}$ indeed. Therefore, for
sufficiently low energy and incoming fermion and negligible
gravitational effect equation (\ref{ACT}) reduces to
\begin{equation}
{\cal S}_f = \int d^2\tilde{x}[i\bar\Psi[\partial\!\!\!/ +
ieA\!\!\!/]\Psi - \frac{1}{4}
e^{-2\Phi(\tilde{x})}F_{\mu\nu}F^{\mu\nu}]. \label{EQ1}
\end{equation}
Here $\gamma^\mu\partial_\mu = \partial\!\!\!/$ and $\gamma^\mu
A_\mu = A\!\!\!/ $ and  $e$ indicates the coupling constant having
the dimension of mass. The Lorentz indices $\mu$ and $\nu$ takes
the values $0$ and $1$ in $(1+1)$ dimensional Minkowsky space
time. The  non dynamical dilatonic background represented by the
field $\Phi$ which is associated here with the electro magnetic
background $F_{\mu\nu}$ has an incredible role in this model
towards  making the coupling constant position dependent. For
later convenience let us now define $\eta^2(x) = e^{2\Phi(x)}$.
The construction of the model is so powerful that it itself
provides a room to take the the effect of anomaly into account
which enters from the one loop correction during the course of
bosonization \cite{MIT, AR, AR1, AR2}. This very anomaly plays a
crucial role to dictate whether there would be disastrous
information loss or it would be free from that disaster. Our
present work does explore that the crucial role of anomaly lies in
the central position in obtaining the information preserving
result when Faddeevian anomaly is considered during the course of
investigation of s-wave scattering of chiral fermion with the
intriguing supportive feature of imposition of chiral constraint
developed by Harada in his seminal work \cite{KH}.

In this article we are intended to present a minute and equitable
analysis of the scattering of chiral fermion in detail in presence
of Faddeevian anomaly proposed by Mitra \cite{PM}. So in the
present situation chiral interaction will take the place of vector
interaction in the same manner as it was done in \cite{MIT, AR2}.
Thus the action in this situation in terms of fermion reads
\begin{equation}
{\cal S}_{CF} = \int d^2\tilde{x}[i\bar\Psi[\partial\!\!\!/ +
ieA\!\!\!/ (1+\gamma_5)]\Psi - \frac{1}{4}
e^{-2\Phi(x)}F_{\mu\nu}F^{\mu\nu}]. \label{SWF}
\end{equation}
In fact, it is the chiral quantum electrodynamics with a dilatonic
background along with the standard electromagnetic background. Our
investigation begins with the bosonized version  of the above
action (\ref{SWF}) to which we now turn. For the shake of
notational convenience we will call $\tilde{x}$ as $x$ in the rest
part of the article.
\section{{\bf Study of s-wave scattering of fermion off dilaton
black-holes with Faddeevian anomaly}} The generating functional of the
above theory  defined by the action (\ref{SWF}) can be written
down as
\begin{equation}
Z[A] = \int d\phi e^{i{\cal S}_{CB}},
\end{equation}
where
\begin{equation}
{\cal S}_{CB}=\int d^2x {\cal L}_{CH},
\end{equation}
with
\begin{equation}
L= \int dx {\cal L}_{CH} = \int dx [
\frac{1}{2}\partial_\mu\phi\partial^\mu\phi + e(\eta^{\mu\nu} -
\epsilon^{\mu\nu})
\partial_\mu\phi A_\nu + \frac{1}{2} e^2[A_0^2 -3A_1^2 - 2 A_1 A_0]
- \frac{1}{4}e^{-2\Phi(x)}F_{\mu\nu}F^{\mu\nu}].\label{NE}
\end{equation}
Equation (\ref{NE}) describes the exact bosonized version of the
fermionic  model descried in (\ref{SWF}) details of which is
presented in the articles \cite{PM, MG, SM}. We will carry out
investigation with the bosonized version of this model (\ref{NE})
because from the article \cite{MIT} we have learned that the study
of chiral fermion in presence of the obstacle posed by the
gravitational anomaly would be possible with the bosonized
version. However, this process of bosonization is crucially
connected with the regularization.  During the course of
integrating out the fermions one gets forced to incorporate the
important one loop correction term in order to get rid of the
divergence appeared in the fermionic determinant. This term may
include different contribution depending on the choice of
regularization. The issue that has to be taken care of with prime
importance during regularization is indeed the maintenance of
physical Lorentz symmetry. In this regard, we should mention that
apparent Lorentz non-covariance not always stands as a hindrance
in order to maintain physical Lorentz invariance \cite{PM, MG,
SM}. Here we choose the Faddeevian type \cite{FADDEEV} of anomaly
developed and studied in the articles \cite{PM, MG, SM}. With this
specialized form of Faddeevian anomaly the bosonized action reads
\begin{equation}
{\cal S}_{CB}=\int
d^2x[\frac{1}{2}\partial_\mu\phi\partial^\mu\phi + e(\eta^{\mu\nu}
- \epsilon^{\mu\nu})
\partial_\mu\phi A_\nu + \frac{1}{2} e^2[A_0^2 -3A_1^2 - 2 A_1 A_0]
- \frac{1}{4}e^{-2\Phi(x)}F_{\mu\nu}F^{\mu\nu}]. \label{BAC}
\end{equation}
Note that there is no Lorentz co-variance at the action level
however in due course we will find that it will not in any way
hamper the physical Lorentz invariance of the theory even in the
presence of dilatonic background. Here
$\epsilon^{01}=-\epsilon_{01}=1$ and the Minkowski metric
$g^{\mu\nu}=diag\left(1,-1\right)$. The indices $\mu$ and $\nu$
takes the values $0$ and $1$ as have already been mentioned in
Sec. II . The advantage of using the bosonized version is that the
anomaly automatically gets incorporated within it. So the tree
level bosonized theory acquires the effect of one loop correction.
Equation (\ref{NE}), apparat from the dilatonic factor was
initially found in \cite{PM} where the author termed it as chiral
Schwinger model with Faddeevian anomaly.

It is now necessary to carry out the Hamiltonian analysis of the
 theory in order to observe the role of dilaton field on the
 theoretical spectra.
What happened to the Lorentz symmetry in the present situation
will also  be explored from our investigation apart from our main
perspective of studying information loss scenario. We need to find
out the theoretical spectrum of this model which necessities to
proceed through the constraint analysis of this theory. From the
standard definition the canonical momenta corresponding to the
chiral boson field $\phi$, the gauge field components $A_0$ and
$A_1$ are computed as follows.
\begin{equation}
\frac{\partial{\cal L}_{CH}}{\partial\dot\phi}= \pi_\phi =
\dot\phi + e(A_0 - A_1),\label{M0}
\end{equation}
\begin{equation}
\frac{\partial{\cal L}_{CH}}{\partial\dot{A}_1}=\pi_1 =
e^{-2\Phi(x)}(\dot{A}_1 - A'_0)=\frac{1}{\eta^2}(\dot A_1 - A_0),
\label{M1}
\end{equation}
\begin{equation}
\frac{\partial{\cal L}_{CH}}{\partial\dot{A}_0}=\pi_0 \approx
0.\label{M2}
\end{equation}
Here $\pi_\phi$ represent the momentum corresponding to the chiral
boson field $\phi$, and  $\pi_1$ and $\pi_0$ stand for the momenta
corresponding to the gauge field components $A_0$ and $A_1$. Using
equations (\ref{M0}), (\ref{M1}) and (\ref{M2}) the canonical
Hamiltonian of the theory is obtained.
\begin{eqnarray}
H_C = \int dx[\frac{1}{2} e^{2\Phi}\pi_1^2
+\frac{1}{2}(\pi_\phi^2+ \phi'^2)+ \pi_1A_0' -e(A_0 -
A_1)(\pi_\phi+\phi') + 2e^2A_1^2].\label{CHAM}
\end{eqnarray}
Note that $\pi_0 \approx 0$ does not contain  any time derivative
of the any field. So it is the primary constraint of the theory.
The preservation of this constraint leads to secondary constraint:
\begin{equation}
G = \pi_1' + e(\pi_\phi + \phi')\approx 0.
\end{equation}
This is generally termed as the Gauss law of the theory.
Therefore, the effective Hamiltonian is given by
\begin{equation}
H_{Ef} = H_C + u\pi_0 + v(\pi_1' + e(\pi_\phi + \phi')).
\label{HEFF}
\end{equation}
Out of these two lagrange multipliers (velocities) $u$ and $v$ the
velocity $v$ is determinable and that comes out to be
\begin{equation}
v = \frac{1}{2}(\pi_\phi + \phi')-e(A_0+A_1),
\end{equation}
but the velocity $u$ remains undetermined at this stage. It just
plays the role of a Lagrange multiplier. The preservation of the
constant $G$ after imposition of the velocity $v$ in the effective
Hamiltonian gives a new constraint.
\begin{equation}
T =(A_0+A_1)\approx 0.
\end{equation}
There is no other constraint in the phase space of the theory. So
the full set of constraints at a glance is
\begin{equation}
\Omega_1 = \pi_0 \approx 0, \label{CN1}
\end{equation}
\begin{equation}
\Omega_2= A_0+A_1\approx 0, \label{CN2}
\end{equation}
\begin{equation}
\Omega_3 = \pi_1' + e(\pi_\phi + \phi')\approx 0. \label{CN3}
\end{equation}
According to the Dirac terminology these constraints are all weak
conditions at this stage. It is found that the Poisson bracket
between $G(x)$ and $G(y)$ gives a non vanishing contribution
\begin{equation} [G(x), G(y)] = 2\delta(x-y)', \label{POIS}
\end{equation}
 which is in sharp contrast to the Poisson bracket in  the case of usual
 chiral Schwinger model \cite{JR} where it was found to render vanishing
contribution. Faddeev initially noticed that anomaly might make
Poisson bracket between $G(x)$ and $G(y)$ nonzero \cite{FADDEEV}
and the constraint became second class itself at that stage and
gauge invariance got lost. He, however, showed that it would be
possible to quantize the theory but in that situation system might
posses extra degrees of freedom. A careful look revels that the
presence of this {\it extra degrees of freedom} causes a problem
in the preservation of information in the scattering problem and
that problem gets eradicated by the imposition of the chiral
constraint which we are going to explored in our present
investigation.

This is a system endowed with three constraints in the phase
space. So ordinary Poisson bracket becomes inadequate and for the
determination of the  phase space structure of this system one
needs the Dirac bracket \cite{DIR} which is defined by
\begin{equation}
[A(x), B(y)]^* = [A(x), B(y)] - \int[A(x), \Omega_i(\eta)]
C^{-1}_{ij}(\eta, z)[\Omega_j(z), B(y)]d\eta dz, \label{DEFD}
\end{equation}
where $C^{-1}_{ij}(x,y)$ can be computed using the following
equation
\begin{equation}
\int C^{-1}_{ij}(x,z) [\Omega_j(z), \Omega_k(y)]dz =\delta(x-y)
\delta_{ik}. \label{INV}
\end{equation}
The system under consideration  described by the Lagrangian
(\ref{NE}), has the following $C_{ij}(x, y)$:
\begin{equation}
C_{ij}(x, y)=\left(\begin{array}{ccc}
 0 &  -\delta(x-y) & 0\\
 \delta(x-y) & 0 & -\delta'(x-y)\\
       0 & -\delta^{\prime}(x-y) & 2e^2\delta'(x-y)
\end{array}\right)
\end{equation}
 The non singular nature of the matrix $C_{ij}$ indicates that it certainly has an
 inverse which is
\begin{eqnarray}
C_{ij}^{-1}(x,y)=\left(\begin{array}{ccc}
-\frac{1}{2e^{2}}\delta'(x-y) & \delta(x-y)& \frac{1}{2e^2} \delta(x-y) \\
       -\delta(x-y)&0 & 0 \\
       \frac{1}{2e^2} \delta(x-y) & 0 &\frac{1}{4e^2}\epsilon(x-y)
\end{array}\right).
\end{eqnarray}
The reduced Hamiltonian is obtained after the imposition of the
constraints(\ref{CN1}),(\ref{CN2}) and (\ref{CN3}) in equation
(\ref{CHAM}) treating these constraints  as strong conditions.
\begin{equation}
{\cal H}_R = \frac{1}{2}(e^{\Phi(x)}\pi_1^2 + \phi'^2) +
\frac{1}{2e^2}\pi_1'^2 + \pi_\phi \phi'-\pi_1A'_1 + 2e^2A_1^2
\label{HAMR}
\end{equation}
Although the field $\Phi$ within the Hamiltonian has space
dependence, it has no explicit time dependence. Hence the
Hamiltonian (\ref{HAMR}) preserves in time. From the definition
(\ref{DEFD}), the Dirac brackets of the fields constituting the
reduced Hamiltonian (\ref{HAMR}) are computed:
\begin{equation}
[A_1(x), A_1(y)]^* = -\frac{1}{{2e^2}}\delta'(x-y), \label{DR10}
\end{equation}
\begin{equation}
[A_1(x), \pi_1(y)]^* = \delta(x-y),\label{DR20}
\end{equation}
\begin{equation}
[A_1(x), \phi(y)]^* = -\frac{1}{2e}\delta(x-y),\label{DR30}
\end{equation}
\begin{equation}
[\pi_1(x), \pi_1(y)]^* = 0, \label{DR40}
\end{equation}
\begin{equation}
[\phi(x), \phi(y)]^* = \frac{1}{4}\epsilon(x-y).\label{DR50}
\end{equation}
Here the superscript $(*)$ indicates the Dirac Bracket. Using the
above Dirac brackets the following first order differential
equations equations of motion are obtained from the
Hamiltonian(\ref{HAMR}).
\begin{equation}
\dot{\pi}_1 = \pi_1'- 4e^2A_1',
\end{equation}
\begin{equation}
\dot{A}_1= e^{2\Phi(x)}\pi_1 - A_1',
\end{equation}
\begin{equation}
\dot{\phi}= -\phi'-\frac{1}{e}\pi_1' +2eA_1.
\end{equation}
After a little algebra the above three equations give the
following theoretical spectra (two second order and one first
order differential equation):
\begin{equation}
(\Box+4e^2e^{2\Phi(x)})A_1 =0, \label{SP1}
\end{equation}
\begin{equation}
(\Box+4e^2e^{2\Phi(x)})\pi_1 =0, \label{SP2}
\end{equation}
\begin{equation}
\partial_{+}[\phi+ \frac{1}{2e}(\dot{A}_1 + A_1')]=0. \label{SP3}
\end{equation}
Here $\partial_{+}= \partial_{0} + \partial_{1}$.  Note that the
the equations of motion are all Lorentz invariant. So Physical
Lorentz invariance is maintained.  The two second order equations
(\ref{SP1}) and (\ref{SP2}) are describing a free massive boson
and its momentum respectively and equation (\ref{SP3}) is
representing a self dual (chiral) boson. The theoretical spectra
therefore contain a massive boson with mass $2ee^{\Phi(x)}$ and an
massless chiral boson \cite{BEL, FJ}. This very chiral boson is
the very {\it extra degrees of freedom} as predicted by Faddeev in
\cite{FADDEEV}. In $(1+ 1)$ dimension, it can be thought of as
chiral fermion. What follows next is the description of the
disaster that was been brought in by this very chiral fermion
towards preservation of information.

Note that because of the presence of the position dependent factor
$\eta^2$ mass of this massive boson varies with the position which
is although astonishing, this position dependence of mass cries
lots of interesting surprises.  The mass in this situation
increases limitlessly in the negative $x^1$ direction because
being inspired by the by the linear dilatonic vacuum of $(1+1)$
dimensional gravity $\Phi(x) = -x^1$ is chosen in the similar
fashion as it was found in the in the earlier articles
\cite{STRO,GARF,GIDD,BANK,CALL,SUS,MIT,AR, AR1,AR2}. So in the
exterior space, where the space like coordinate $x^1 \to\infty$,
the coupling $\eta^2(x)$ vanishes exponentially \cite{STRO}.
However, when $x^1 \to -\infty$ the coupling constant will diverge
that can be considered as an infinite throat in the interior of
certain magnetically charged black-hole. Therefore, the massive
particle will not be able to travel an arbitrary distance so it
will be return back and any finite energy contribution must
therefore be totally come back with unit probability. So an
observer at $x^1 \to \infty$ will be able to retrieve all the
information of the massive field. More rigorously, we can say that
because of the position dependent factor $\eta$ within the mass
term it will start journey with the vanishing value near the mouth
but as it goes into the throat region mass will increases
limitlessly. Since in $(1+1)$ a massless scalar can be thought of
as a massless fermion so a fermion which will proceed into the
black-hole will not be able to advance an arbitrary long distance
and will certainly be reflected back with a unit probability.
Therefore, the massive boson will not create any problem to
maintain preservation of information. However the mass less chiral
fermion field will be able to travel smoothly within the
black-hole without facing any obstruction. So the observer at $x^1
\to \infty$ will not find this massless chiral field to return
back. Hence the total information corresponding to the mass lees
chiral fermion will be lost. Even though it is a puzzling outcome
with this setting and it goes against the Hawking's revised
proposal related to information preserving scenario \cite{HAW1} it
is unavoidable at this stage. Although the anomaly considered here
is Faddeevian type, the result has gone against the preservation
of information. So a natural question may be initiated: how does
the information preserving result for s-wave scattering of chiral
fermion off dilaton black-hole was obtained in the article
\cite{AR1}? In the following Sec. IV, we will proceed to take an
attempt to uncover where the crux of the fact is lying hidden.

\section{A remedy to the disaster by the imposition of chiral constraint}
From the  discussion of the earlier Sec. III, it has been
confirmed that the presence of the very chiral fermion (\ref{SP3})
is in the source of all disaster towards non preservation of
information. So this Sec. of the article will be devoted to
eliminate this chiral fermion from the spectra  with a suitable
standard physical technique keeping our intense focus on the fact
to recover the preservation of information.

To search for a remedy let us see how does the remarkable
technique of imposition of a chiral constraint presented by Harada
in the article \cite{KH} can be brought into the service? It is
known from quite a long ago that  Harada in his seminal work
\cite{KH} introduced this powerful technique to develop the gauged
version of Flourinini-Jackiw type of chiral boson \cite{FJ} from
the usual Chiral Scwinger model with Jackiw-Rajaraman type of
anomaly. We are now going to exploit the functionally active power
of that fascinating technique to protect this model from this
disastrous information loss problem.

At this stage, we will therefore proceed to impose a  chiral
constrain in the model described in equation (\ref{NE}). During
the course of imposition of chiral constraint, we will not take
into account the kinetic part of electro magnetic background
accompanied with the dilaton field, because it remains unaffected
during the course of imposition of chiral constraint. We indeed
take that part into account in due course when we will proceed to
determine the theoretical spectrum of this interacting model.
Apart from the kinetic part of electromagnetic field entangled
with non dynamical dilatonic background the model reads
\begin{equation}
{\cal L}_{CHB} = \frac{1}{2}\partial_\mu\phi\partial^\mu\phi +
e(\eta^{\mu\nu} - \epsilon^{\mu\nu})
\partial_\mu\phi A_\nu + \frac{1}{2} e^2[A_0^2 -3A_1^2 - 2 A_1 A_0].
\end{equation}
In order to  impose the chiral constraint we need to calculate the
momentum corresponding to the field $\phi$. From the standard
definition the momentum corresponding to the field $\phi$ comes
out as
\begin{equation}
\frac{\partial{\cal L}_{CHB}}{\partial\dot\phi} = \pi_\phi =
\dot\phi + e(A_0 - A_1).
\end{equation}
The Hamiltonian of this system as usual is obtained through the
Legendre transformation:
\begin{equation}
H_B = \int d^2x [\pi_\phi\dot\phi - {\cal L}_{CHB}]. \label{LEGT}
\end{equation}
Explicitly, the Hamiltonian density of the system as obtained from
the Legendre transformation (\ref{LEGT}), is
\begin{eqnarray}
{\cal H}_B = \frac{1}{2}[\pi_\phi - e(A_0 - A_1)]^2 +
\frac{1}{2}\phi'^2 - 2e\phi'(A_0 - A_1) - \frac{1}{2}e^2(A_0^2 -2
A_0A_1 -3A_1^2). \label{CHBH}
\end{eqnarray}
We are now in a position to impose the chiral constraint in the
phase space of the system described by the Hamiltonian
(\ref{CHBH}). The
 chiral constraint which has been brought here into the service is
\begin{equation} \Omega(x) = \pi_\phi(x) -
\phi'(x) \approx 0, \label{CHC}
\end{equation} which has the following Poisson bracket with itself
\begin{equation}
[\Omega(x), \Omega(y)] = -2\delta'(x-y). \end{equation}
 Therefore,
this chiral constraint itself is  second  class in nature.
 After
imposing the constraint $\Omega(x) \approx  0$, into the phase
space the generating functional of the corresponding theory reads
\begin{eqnarray}
Z_{CH} &=& \int d\phi d\pi_\phi \delta(\pi_\phi - \phi')
\sqrt{det[\Omega, \Omega]}e^{ i\int d^2x(\pi_\phi\dot\phi - {\cal
H}_B)},
\nonumber \\
&=&\int d\phi e^{i\int d^2x{\cal L}_{CH}}, \end{eqnarray} with
\begin{equation}
{\cal L}_{CH} = \dot\phi\phi' -\phi'^2 + 2e(A_0 - A_1)\phi' - 2e^2
A_1^2.\label{LCH}
\end{equation}
This is the gauged Lagrangian density for Floreanini-Jackiw type
chiral boson apart from the electromagnetic background and it is
obtained from the bosonized version of chiral Schwinger with
Faddeevian type anomaly \cite{PM} after implementing the chiral
constraint in its phase space in the same fashion  as it was found
to be implemented by Harada in \cite{KH} in order to formulate a
model of the gauged chiral boson from the usual chiral Schwinger
model with one parameter class of anomaly introduced  by Jackiw
and Rajaraman \cite{JR}. The constraint analysis and the
determination of the phase space structure corresponding to this
model will almost alike to the work presented in \cite{MG}. The
only difference that will appear is due to the presence of the non
dynamical dilatonic factor entangled with the kinetic term of the
electromagnetic background. Let us now take into account the
electromagnetic background which is entangled with dilaton field
crucially with the starting Lagrangian. The resulting model now
takes the following form
\begin{eqnarray}
{\cal L}_{CHC} = \dot\phi\phi' -\phi'^2 + 2e(A_0 - A_1)\phi' -
2e^2 A_1^2 + \frac{1}{2} e^{-2\Phi(x)}F_{01}^2. \label {LCHC}
\end{eqnarray}
Here $\phi$ represents a chiral boson field. The tree level
bosonized theory has acquired the effect of anomaly in this
situation too \cite{MG, SM}.

We will now proceed with the Hamiltonian analysis of the theory to
observe the role of dilaton field on the theoretical spectrum in
the present situation, i.e. after the imposition of the chiral
constraint $\Omega(x)\approx0$ . To determine the theoretical
spectrum of  this interacting model in the present situation we
have to use the constrained dynamics developed by Dirac
\cite{DIR}. So we need to calculate the momentum corresponding to
the fields with which the theory is constituted. From the standard
definition the canonical momenta corresponding to the chiral boson
field $\phi$, the gauge field $A_0$ and $A_1$ are obtained:
\begin{equation}
\frac{\partial{\cal L}_{CHC}}{\partial\dot\phi} = \pi_\phi =
\phi',\label{MO1}
\end{equation}
\begin{equation}
\frac{\partial{\cal L}_{CHC}}{\partial\dot{A}_0}= \pi_0 \approx
0,\label{MO2}
\end{equation}
\begin{equation}
\frac{\partial{\cal L}_{CHC}}{\partial\dot{A}_1}= \pi_1 =
e^{-2\phi(x)}(\dot A_1 - A_0')=\frac{1}{\eta^2}(\dot A_1 -
A_0).\label{MO3}
\end{equation}
Here $\pi_\phi$, $\pi_0$ and $\pi_1$ are the momenta corresponding
to the field $\phi$, $A_0$ and $A_1$. Note that there are two
primary constraint in the phase space of the theory:
\begin{equation}
\tilde\Omega_1 = \pi_0 \approx 0, \label{PCON1}
\end{equation}
\begin{equation}
\tilde\Omega_2 = \pi_1 + e\phi \approx  0,\label{PCON2}
\end{equation}
Using the  equations (\ref{MO1}),(\ref{MO2})  and (\ref{MO3}) it
is straightforward to obtain the canonical Hamiltonian through a
Legendre transformation:
\begin{equation}
{\cal H}_C =\pi_\phi \dot{\phi} + \pi_1 \dot{A}_1- {\cal L}.
\end{equation}
The  above transformation yields
\begin{equation}
H_C =\int dx[\frac{1}{2} e^{2\Phi}\pi_1^2 + \pi_1A_0' + \phi'^2
-2e(A_0 - A_1)\phi' + e^2A_1^2].\label{CBHAM}
\end{equation}

The Hamiltonian acquires an explicit space dependence through
$\Phi(x)$, however it has no time dependence so it is preserved in
time in this situation too. Equation (\ref{PCON1}) and
(\ref{PCON2}) are the primary constraints of the theory. So it
necessities to deal with the effective Hamiltonian:
\begin{equation}
H_{EFF} = H_C + {\tilde u}\pi_0 + {\tilde v}(\pi_\phi - \phi').
\label{EF}
\end{equation}
Here ${\tilde u}$ and ${\tilde v}$ are two arbitrary Lagrange
multipliers. The consistency of the  theory  requires the
preservation of the constraint (\ref{PCON1}) and (\ref{PCON2}) in
time. The preservation of the constraint (\ref{PCON1}) leads to a
new constraint which is the Gauss law of the theory.
\begin{equation}
\tilde\Omega_4(x) = (\pi_1' + 2e\phi')(x)\approx 0. \label{GAUS}
\end{equation}
The preservation of the constraint (\ref{GAUS}) initially does not
give rise to any new constraint  since it fixes the velocity
${\tilde v}$ which comes out to be
\begin{equation}
{\tilde v} = \phi' - e(A_0 - A_1). \label{VEL}
\end{equation}
After substituting the velocity ${\tilde v}$ in (\ref{EF}) the
consistency of the theory leads to a new constraint
\begin{equation}
\tilde\Omega_3= (A_0 + A_1)'\approx 0.\label{FINC}
\end{equation}
We do not get any new constraints from the preservation of
(\ref{FINC}). So, we find that the phase space of the theory is
endowed with the  following four constraints:
\begin{equation}
\tilde\Omega_1 = \pi_0 \approx 0, \label{CON1}
\end{equation}
\begin{equation}
\tilde\Omega_2 = \pi_1 + e\phi \approx  0,\label{CON2}
\end{equation}
\begin{equation}
\tilde\Omega_3 = (A_0 + A_1) \approx 0,\label{CON3}
\end{equation}
\begin{equation}
\tilde\Omega_4 = \pi_\phi - \phi' \approx 0. \label{CON4}
\end{equation}
The four constraints (\ref{CON1}), (\ref{CON2}), (\ref{CON3}) and
(\ref{CON4}) form a second class set and all are weak condition up
to this stage. If we impose these constraints  into the canonical
Hamiltonian (\ref{CBHAM}) treating these as strong conditions, the
canonical Hamiltonian will be simplified into
\begin{equation}
H_{CR} = \int dx^1[\frac{1}{2}e^{2\Phi(x)}\pi_1^2 +
\frac{1}{{4e_2}} \pi_1'^2 - \pi_1'A_1 + 2e^2 A_1^2]. \label{RHAM}
\end{equation}
The Hamiltonian $H_{CR}$ lying in equation (\ref{RHAM}), is
generally known as reduced Hamiltonian. According to Dirac's
formalism of constrained dynamics, Poisson brackets become
inadequate for this reduced Hamiltonian \cite{DIR}. This reduced
Hamiltonian however remains consistent with the Dirac brackets
which has already been defined in equation (\ref{DEFD})

For this constrained system the matrix constituted with the
Poisson brackets among the constraints themselves is
\begin{equation}
C_{ij}=\left(\begin{array}{cccc}
0 & 0 &-\delta(x-y) & 0 \\
       0 & 0 & -\delta(x-y) & 2e\delta(x-y)\\
       \delta(x-y) & e\delta(x-y) & 0 & 0 \\
       0 & 2e\delta(x-y)  & 0 & -2\delta'(x-y)
\end{array}\right)
\end{equation}
Since $C_{ij}$ in this situation too remains nonsingular it is
invertible.  Using equation (\ref{INV}) the inverse matrix
$C^{-1}_{ij}$ is computed and it is given by
\begin{eqnarray}
C_{ij}^{-1}=\left(\begin{array}{cccc}
\frac{2}{e^2}\delta'(x-y) & \frac{1}{2e^2}\delta'(x-y) & \delta(x-y)) & -\frac{1}{2e} \delta(x-y) \\
       -\frac{1}{2e}\delta'(x-y) & -\frac{1}{2e}\delta'(x-y) & 0 & \frac{1}{2e}\delta(x-y) \\
       - \delta(x-y)& 0 & 0 & 0 \\
       \frac{1}{2e}\delta(x-y) & -\frac{1}{2e}\delta(x-y) & 0 & 0
\end{array}\right).
\end{eqnarray}

From the definition of the Dirac brackets (\ref{DEFD}), we can
compute the Dirac brackets between  the fields $A_1$ and $\pi_1$
those which have been describing the reduced Hamiltonian $H_{CR}$.
These are required to obtain the theoretical spectra in the
present situation.
\begin{equation}
[A_1(x), A_1(y)]^* = -\frac{1}{{2e^2}}\delta'(x-y), \label{DR1}
\end{equation}
\begin{equation}
[A_1(x), \pi_1(y)]^* = \delta(x-y),\label{DR2}
\end{equation}
\begin{equation}
[\pi_1(x), \pi_1(y)]^* = 0 \label{DR3}
\end{equation}
The superscript (*) is representing the Dirac bracket as usual.
Using the Dirac brackets (\ref{DR1}), (\ref{DR2}) and (\ref{DR3}),
we obtain the following two first order equations of motion from
the reduced Hamiltonian (\ref{RHAM}):
\begin{equation}
\dot A_1 = e^{2\Phi(x)}\pi_1 - A_1', \label{EQM1}
\end{equation}
\begin{equation}
\dot \pi_1 = \pi_1' - A'_1. \label{EQM2}
\end{equation}
We find that both the field $A_1$ and $\pi_1$ and  satisfy  Klein
Gordon type equation:
\begin{equation}
(\Box + 4 e^{2\Phi(x)}e^2)A_1 = 0 \label{SPEC0},
\end{equation}
\begin{equation}
(\Box + 4 e^{2\Phi(x)}e^2)\pi_1 = 0 \label{SPEC1}.
\end{equation}
So the system contains only a massive boson with mass
$2ee^{\Phi}(x)$. The equation (\ref{SPEC1}) represents the
equation of motion of momentum corresponding to the field $A_1$.
Note that the fields $A_1$ and $\pi_1$ satisfy canonical Dirac
bracket in the constrained sub space.
It is amazing to note that there is no mass less chiral boson as
was found in the previous situation according to the prediction
made by Faddeev in the article \cite{FADDEEV}. The very presence
of chiral boson in the preceding Sec. III are found to be in the
root of the disastrous information loss because it was capable of
travelling an arbitrary distance without getting any hindrance
since the dilatonic background failed to pose any obstruction to
the free chiral fermion and the observer $x^1 \to \infty$ did not
find it with a backward journey. So information loss was
inevitable there. But the present situation appears with a
remarkable difference so far constraint structure is concerned and
that in turn leads to a welcome change in the theoretical
spectrum. The disturbing extra degrees of freedom in the form of
chiral boson which can be thought of in terms of a chiral fermion
in (1+1) dimension miraculously gets disappeared from the
theoretical spectrum and all credit goes to that important
technique of imposition of chiral constraint due to Harada
\cite{KH}. Since there is no chiral fermion in the theoretical
spectrum, there will be no scope of information loss and a unitary
s-matrix can be constructed for this situation. This results
reminds us the s-wave scattering of Dirac fermion too as presented
in the article \cite{STRO}. Thus the scattering of chiral fermion
with Faddeevian type of anomaly along with the imposition of the
chiral constraint is found to be free from the dangerous
information loss problem. This result is completely opposite to
the observation of the authors of the article \cite{MIT} where the
authors considered the same problem with different type of
anomaly.

\section{\bf{Discussion and conclusion}}
The analysis available in article \cite{MIT} and the present
analysis differs only in the choice of taking the anomaly into
account which enters during the course of bosonization through a
one loop correction, but that has brought a  remarkable change in
the information loss scenario. Note that the model used in the
article \cite{MIT} was also generated by the imposition of the
same chiral constraint by Harada developed  in the article
\cite{KH}, but the authors of the article \cite{MIT} obtained a
result which was not in agreement with the revised suggestion of
Hawking \cite{HAW1}. This present result is consistent with the
standard belief as well as it is in agreement with the Hawking's
revised suggestion \cite{HAW1}. Therefore although the role of
Fadeevian anomaly is crucial, the importance of the technique of
imposition of chiral constraint is indeed instrumental along with
that. On other hand, we would have got information preserving
result in Sec. III, where imposition of this very chiral
constraint (\ref{CHC}) was not implemented.

Apart from studying information related problem through the s-wave
scattering of fermion off dilaton black-hole there are several
approaches to study this information preservation related question
\cite{VEG, CHEN, SUDIP, AHN}. The information preserving result as
we have obtained hare also agrees with the recent result obtained
in \cite{VEG, CHEN}. In order to get the information preserving
result in the present situation we exploit the regularization
ambiguity that crucially alters the anomaly structure. Along with
that a very useful technique of imposition of chiral constraint
\cite{KH} also has been employed. And these two are equally
important for holding the information preserving result. These are
very often found in quantum electro dynamics and chiral quantum
electrodynamics. A remarkable instance in this context is the
removal of the the long suffering of chiral generation of
Schwinger model \cite{HAG, SCH} from non-unitarity problem  by
Jackiw and Rajaraman in their seminal work \cite{JR} inviting the
anomaly into the model. We, therefore, conclude that with the
support of the standard technique of imposition of chiral
constraint due to Harada \cite{KH}, the anomaly that enterd into
the theory as a one loop correction during bosonization has been
found to play a crucial role in dictating whether the scattering
of chiral fermion off magnetically charged dilaton black-hole will
face the information loss problem or it will get averted from that
danger. The main focus of this article is to establish that
Faddeevian anomaly can save the model from dangerous information
loss problem with the incredible support of the standard technique
of imposition of chiral constraint due to Harada \cite{KH}.
Surprisingly, it would not be possible for the standard anomaly
considered by Jackiw-Rajaraman to protect the s-wave scattering of
chiral fermion off dilaton black-holes from the disastrous
information loss problem even if the support of the standard
technique of imposition of chiral constraint would be taken there
as has been found in the important study of Mitra \cite{MIT}.
Therefore, so far preservation of information is concerned through
s-wave scattering of dilaton black-hole the Faddevian type of
anomaly \cite{FADDEEV} scores over the Jackiw-Rajaraman type of
anomaly \cite{JR}.

We have seen various examples with the crucial dependence of
information loss with anomaly via the exploitation of ambiguity in
the articles presented earlier in different times \cite{STRO, MIT,
AR, AR1, AR2}. The present one has enlarged this list to confirm
more convincingly the crucial role of anomaly towards the
information preserving scenario with the support of the standard
and powerful technique of imposition of chiral constraint
\cite{KH}. The crucial role of exploitation of ambiguity in the
regularization has been found in different perspective also. It is
very common in connection with the confinement scenario of fermion
in (1+1) dimensional electrodynamics and chiral electrodynamics
\cite{AR,AR1,MG,ARPM,ARFR, ARAN, ARAN1}. With the usual anomaly
the fermion was found to be unconfined where as with the
Faddeevian type of anomaly it was found to remain confined
\cite{MG}. In that case also the supportive role of this powerful
technique of imposition of chiral constraint developed by Harada
\cite{KH} can be observed if a careful look is projected \cite{MG,
ARFR}. However, it is fair to admit that hitherto there is no
known direct relation between the confinement of fermion with the
anomaly.

We should mention that in the article \cite{AR2}, we considered
the generalized Faddeevian anomaly, however in our present article
we have considered  the specialized Mitra type Faddeevian anomaly
which was presented quite a long ago in the article \cite{MIT}.
The whole algorithm can be carried out with the generalized
Fadeevian anomaly too which will provide a prototype description
but that would render no new information.

Besides, we must mention here that the information loss problem as
faced in the article \cite{MIT} was much more severe than this
one, since we have shown that although the use of the fascinating
technique of imposition of chiral constraint developed  by Harada
\cite{KH} can save this model with Faddeevian anomaly from the
danger of information loss, hitherto there is no known standard
formalism to save the model studied in the article \cite{MIT} from
the disastrous information loss problem. Few literatures are
available where attempts have been made to protect different
models to save them from this disastrous information loss problem
\cite{AR, AR2}. In the article \cite{AR2}, a very effective
nonstandard technique has been used where the service of
unparticle \cite{GEOR, GEOR1} was brought into action to make the
scattering process information preserving. However,  none of the
techniques are found effective to save the model studied in the
article \cite{MIT} from this disastrous information loss.  But one
might be hopeful to save this model too from this problem.
Intensive investigation is needed for that purpose indeed.

We must mention here that the general study of this problem
\cite{HOOFT} is accompanied with lots of technical complexities
and computational difficulties. Therefore, most of the authors
including us prefer to deal with this problem with this less
complicated but well formulated interacting model presented in
\cite{STRO}.

\noindent {\bf Acknowledgments} I would like to extend my sincere
thank to the Director, Saha Institute of Nuclear Physics, Kolkata,
for providing computer and library facilities of the Institute.

\end{document}